\documentclass[twocolumn,showpacs,preprintnumbers,amsmath,amssymb]{revtex4}

\usepackage{graphicx}
\usepackage{dcolumn}
\usepackage{bm}

\begin{document}


\title{Analytical solution of $t\bar{t}$ dilepton equations}

\author{Lars Sonnenschein}

\email{Lars.Sonnenschein@cern.ch}

\affiliation{RWTH Aachen University, III. Physikalisches Institut A, 52056 Aachen, Germany}

\date{\today} 

\begin{abstract}
The top quark antiquark production system in the dilepton decay channel is described by a set 
of equations which is nonlinear in the unknown neutrino momenta. Its most precise and least 
time consuming solution is of major importance for measurements of top quark properties like 
the top quark mass and $t\bar{t}$ spin correlations. 
The initial system of equations can be transformed into two polynomial equations with two 
unknowns by means of elementary algebraic operations. These two polynomials of multidegree
two can be reduced to one univariate polynomial of degree four by means of resultants.
The obtained quartic equation is solved analytically.

\end{abstract}

\pacs{PACS29.85.+C}
\keywords{top dilepton, kinematics, system of equations, solution}
\maketitle

\section{Introduction}

\noindent
In 1992, Dalitz and Goldstein published a numerical method based on geometrical considerations to solve the system of equations describing the kinematics of the $t\bar{t}$ decay in the dilepton channel \cite{Dalitz1992}\cite{Goldstein1992}.
In 2004 an approximation of the system of equations - assuming that the transverse momentum
of the $t\bar{t}$ system can be neglected - has been solved analytically \cite{Homola2004} by means of computer algebra software such as \cite{Maple2004}.
Meanwhile the transverse momentum constraint has been omitted
while the solution is still derived by means of computer algebra and 
its accuracy does not reach real precision \cite{Homola2005}.
In 2005, the system of equations could be solved algebraically to real precision 
free of any singularity \cite{ttdilepsol2005}.
The analytical solution introduced here is based on a new Ansatz which minimises the amount 
of intermediate steps to derive the solution. This approach makes the need of computer algebra
superfluous. In addition it provides more transparency and control over singularities 
which are intrinsic to the analytical solution. 
Further the accuracy achieved is - as already in the 
algebraic approach \cite{ttdilepsol2005} - of real precision.
Important improvements in terms of robustness, code volume and time consumption
with respect to the algebraic approach make this method more convenient
for applications in practice. Other solution methods can compare their performance
to the reference method described here. 
It should be mentioned that different approachs leading to analytical solutions, 
without giving a complete algebraic derivation and without rigorous
discussion of reducible and irreducible singularities exist in the 
literature \cite{Zhou1998} \cite{Sjolin2003}.

In the next section the system of $t\bar{t}$ dilepton equations
is introduced, followed by the derivation of the analytical solution
including a rigorous discussion of the reducible and irreducible singularities
of the analytical solution. 
Subsequently, the performance of the method is elaborated.

\section{$t\bar{t}$ dilepton kinematics}

\noindent
The system of equations describing the kinematics of $t\bar{t}$
dilepton events can be expressed by the two linear and six non linear
equations
\begin{eqnarray} 
\label{initialequations} 
\nonumber 
E_x\!\!\!\!\!\!/ \;\;& = & p_{\nu_x} + p_{\bar{\nu}_x} , \\ \nonumber
E_y\!\!\!\!\!\!/ \;\;& = & p_{\nu_y} + p_{\bar{\nu}_y} , \\ \nonumber
E_{\nu}^2 & = & m_{\nu}^2 + p_{\nu_x}^2+p_{\nu_y}^2+p_{\nu_z}^2 , \\ \nonumber 
E_{\bar{\nu}}^2 & = & m_{\bar{\nu}}^2 + p_{\bar{\nu}_x}^2+p_{\bar{\nu}_y}^2+p_{\bar{\nu}_z}^2 , \\ \nonumber 
m_{W^+}^2 \!\! & = & (E_{\ell^+}+E_{\nu})^2-(p_{\ell^+_x}+p_{\nu_x})^2 , \\ \nonumber
 & & -(p_{\ell^+_y}+p_{\nu_y})^2-(p_{\ell^+_z}+p_{\nu_z})^2 , \\
m_{W^-}^2 \!\! & = & (E_{\ell^-}+E_{\bar{\nu}})^2-(p_{\ell^-_x}+p_{\bar{\nu}_x})^2 , \\ \nonumber
 & & -(p_{\ell^-_y}+p_{\bar{\nu}_y})^2-(p_{\ell^-_z}+p_{\bar{\nu}_z})^2 , \\ \nonumber
m_t^2 \; & = & (E_b+E_{\ell^+}+E_{\nu})^2-(p_{b_x}+p_{\ell^+_x}+p_{\nu_x})^2 , \\ \nonumber
 & & -(p_{b_y}+p_{\ell^+_y}+p_{\nu_y})^2-(p_{b_z}+p_{\ell^+_z}+p_{\nu_z})^2 , \\ \nonumber
m_{\bar{t}}^2 \; & = & (E_{\bar{b}}+E_{\ell^-}+E_{\bar{\nu}})^2-(p_{\bar{b}_x}+p_{\ell^-_x}+p_{\bar{\nu}_x})^2 , \\ \nonumber
 & & -(p_{\bar{b}_y}+p_{\ell^-_y}+p_{\bar{\nu}_y})^2-(p_{\bar{b}_z}+p_{\ell^-_z}+p_{\bar{\nu}_z})^2 . 
\end{eqnarray}
The $z$-axis is here assumed to be parallel orientated to the beam axis while 
the $x$- and $y$-coordinates span the transverse plane.
The first two equations relate the projection of the missing transverse energy
onto one of the transverse axes ($x$ or $y$) to the sum of the neutrino and 
antineutrino momentum components belonging to the same projection.
The next two equations relate the energy of the neutrino and antineutrino
with their momenta. 
Finally four non linear equations describe the $W$ boson and top quark
(antiquark) mass constraints by relating the invariant masses to the
energy and momenta of their decay particles via relativistic 4-vector
arithmetics.

\section{Analytical solution}
\label{anasol}

\noindent
The system of equations (\ref{initialequations}) can be subdivided in two entangled 
sets of equations. One set of equations,
describing the $t\rightarrow bW^+\rightarrow b\ell^+\nu_{\ell}$ parton branch 
of the event, depends on $p_{\nu_z}$ while the other pair of equations, 
describing the $\bar{t}\rightarrow \bar{b}W^-\rightarrow \bar{b}\ell^-\bar{\nu}_{\ell}$ parton branch
of the event, depends on $p_{\bar{\nu}_z}$.

The equation describing the invariance of the $W$ boson mass can be expressed
in the following way
\begin{eqnarray} \nonumber
m_{W^+}^2 & = & (E_{\ell^+}+E_{\nu})^2-(\vec{p_{\ell^+}}+\vec{p_{\nu}})^2 \\ \nonumber
 & = & E_{\ell^+}^2+2E_{\ell^+}E_{\nu}+E_{\nu}^2-\vec{p_{\ell^+}}^2-2\vec{p_{\ell^+}}\vec{p_{\nu}}-\vec{p_{\nu}}^2 \\ 
 & = & m_{\ell^+}^2+m_{\nu}^2+2E_{\ell^+}E_{\nu}-2\vec{p_{\ell^+}}\vec{p_{\nu}} \\ \nonumber
\end{eqnarray}
which can be rewritten as
\begin{eqnarray} \label{wmasseq}
  E_{\nu} & = & \frac{m_{W^+}^2-m_{\ell^+}^2-m_{\nu}^2+2\vec{p_{\ell^+}}\vec{p_{\nu}}}{2E_{\ell^+}} .
\end{eqnarray}
The equation describing the invariance of the top quark mass can be transformed in the same way
leading to 
\begin{eqnarray} \label{tmasseq}
 E_{\nu} & \!\!\! = \!\!\! & \frac{m_t^2\!\!-\!m_b^2\!\!-\!\!m_{\ell^+}^2\!\!-\!m_{\nu}^2\!\!\!-\!\!2E_bE_{\ell^+}\!\!\!+\!\!2\vec{p_b}\vec{p_{\ell^+}}\!\!+\!\!2(\vec{p_b}\!\!+\!\!\vec{p_{\ell^+}}\!)\vec{p_{\nu}}}{2(E_b+E_{\ell^+})} \;\;\;\;
\end{eqnarray}
where additional terms emerge due to the fact that quantities which depended in equation (\ref{wmasseq}) 
only on the lepton depend now also on the $b$ quark.
Next the unknown $E_{\nu}$ can be eliminated by subtracting equation (\ref{tmasseq}) from
(\ref{wmasseq}), leading to an equation of the form
\begin{eqnarray} \label{linpnuz}
  0 & = & a_1 + a_2p_{\nu_x} + a_3p_{\nu_y} + a_4p_{\nu_z}
\end{eqnarray}
where the coefficients $a$ are constants given in the appendix. This equation is linear in the 
three neutrino momentum components. 
Since the unknown $p_{\nu_z}$ does only appear in the top quark parton branch
it is mandatory to eliminate this variable with a linear independent equation
of the top quark parton branch to obtain finally together with the equations 
of the antitop quark branch two equations of the two unknowns $p_{\nu_x}$ and $p_{\nu_y}$.

\begin{figure}[b]
\hspace*{0ex}\includegraphics[width=9.1cm]{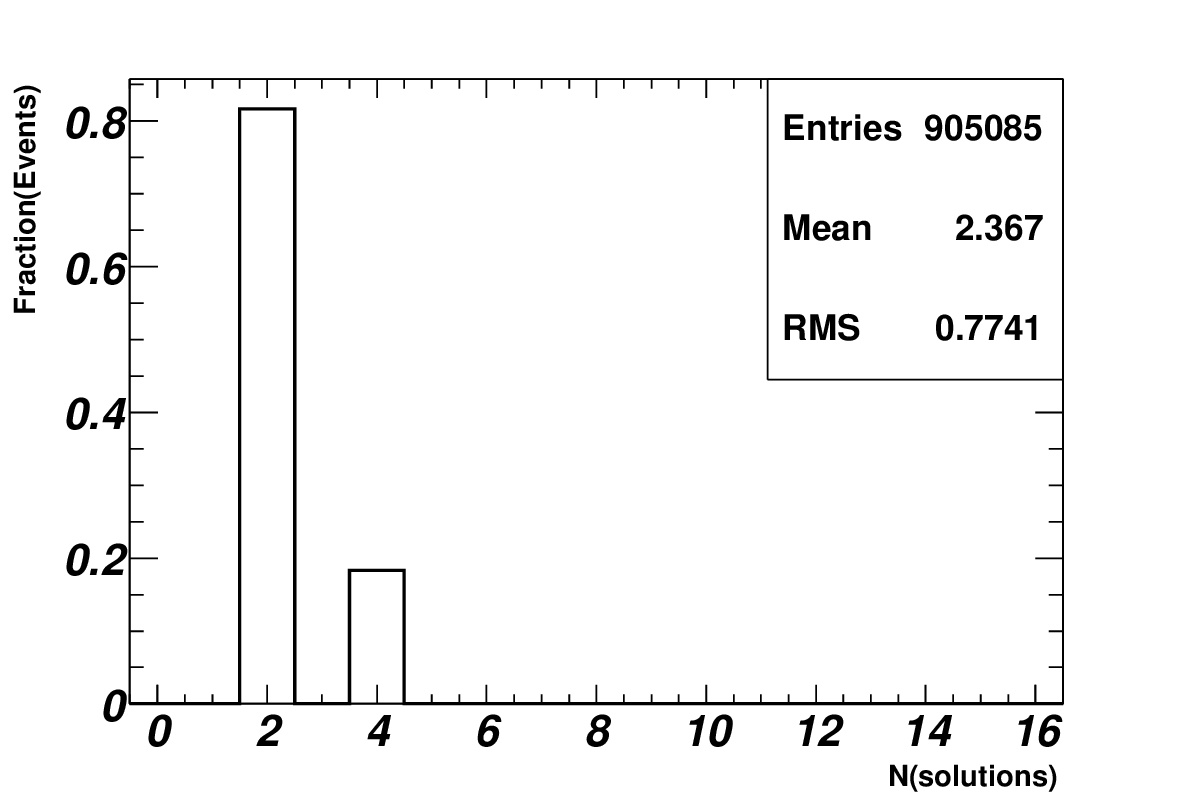}
\caption{\label{Nsol} Number of solutions per event for particles before any radiation.
$t$ quark and $W$ boson masses are assumed to be known exactly.}
\end{figure}

To eliminate the unknown $p_{\nu_z}$
it is straight forward to use equation (\ref{wmasseq}) (for convenience 
multiplied by the denominator $2E_{\ell^+}$). The neutrino energy $E_{\nu}$
can be expressed in terms of the three neutrino momenta components in 
substituting it with the third equation of (\ref{initialequations}). To obtain
a polynomial equation the squared of this equation is
being considered in the following. 
Its longitudinal neutrino momenta 
can be substituted by equation (\ref{linpnuz}), solved to $p_{\nu_z}$.
The resulting equation of the form
\begin{eqnarray} \label{pnuxypol}
  0 \! & \! = \! & \! c_{22}\!+\!c_{21}p_{\nu_x}\!\!+\!c_{11}p_{\nu_y}\!\!+\!c_{20}p_{\nu_x}^2\!\!+\!c_{10}p_{\nu_x}p_{\nu_y}\!\!+\!c_{00}p_{\nu_y}^2
\end{eqnarray}
is a multivariate polynomial of multidegree two which depends only on the transverse 
neutrino momenta $p_{\nu_x}$ and $p_{\nu_y}$.
The coefficients are again constants which can be expressed in terms of the former derived constants
$a$ and are given in the appendix.

In the same way can be proceeded for the equations describing the antitop quark parton branch.
The equivalent of equation (\ref{linpnuz}) reads
\begin{eqnarray} \label{linpnubarz}
  0 & = & b_1 + b_2p_{\bar{\nu}_x} + b_3p_{\bar{\nu}_y} + b_4p_{\bar{\nu}_z}
\end{eqnarray}
and the counter part of polynomial (\ref{pnuxypol}) can be written as
\begin{eqnarray} \label{pnubarxypol}
  0 \! & \! = \! & \! d'_{22}\!+\!d'_{21}p_{\bar{\nu}_x}\!\!+\!d'_{11}p_{\bar{\nu}_y}\!\!+\!d'_{20}p_{\bar{\nu}_x}^2\!\!+\!d'_{10}p_{\bar{\nu}_x}p'_{\bar{\nu}_y}\!\!+\!d'_{00}p_{\bar{\nu}_y}^2 \, .
\end{eqnarray}

The two equations linear in the three (anti-)neutrino momenta (\ref{linpnuz}) and (\ref{linpnubarz}) build the minimal Ansatz
used here. In contrast the Ansatz made in \cite{Homola2004}, \cite{Homola2005}
is based on two equations linear in the four unknowns
$p_{\bar{\nu}_x}$, $p_{\bar{\nu}_y}$, $p_{\bar{\nu}_z}$, $p_{\nu_z}$. 

To reduce equations (\ref{pnuxypol}) and (\ref{pnubarxypol}) to two polynomial equations of two unknowns
the transverse antineutrino momenta of equation (\ref{pnubarxypol}) can be expressed by the transverse neutrino momenta
with help of the missing transverse energy relations of the system of equations (\ref{initialequations}).
Since these relations are linear in the neutrino and antineutrino momenta the substitution 
leads again to a polynomial of the form 
\begin{eqnarray} \label{pnuxypol2}
  0 \! & \! = \! & \! d_{22}\!+\!d_{21}p_{\nu_x}\!\!+\!d_{11}p_{\nu_y}\!\!+\!d_{20}p_{\nu_x}^2\!\!+\!d_{10}p_{\nu_x}p_{\nu_y}\!\!+\!d_{00}p_{\nu_y}^2
\end{eqnarray}
with multidegree two whose coefficients are given in the appendix.
To solve these two polynomials without loss of generality to $p_{\nu_x}$ the resultant with respect 
to the neutrino momentum $p_{\nu_y}$ is computed as follows. The coefficients and monomials of the 
two polynomials (\ref{pnuxypol}) and (\ref{pnuxypol2}) are rewritten in such a way that they are 
ordered in powers of $p_{\nu_y}$ like
\begin{eqnarray} \label{cpoly}
  c & = & c_0p_{\nu_y}^2+c_1p_{\nu_y}+c_2 , \\ 
  d & = & d_0p_{\nu_y}^2+d_1p_{\nu_y}+d_2 \label{dpoly}
\end{eqnarray}
where $c$ and $d$ are polynomials of the remaining unknowns $p_{\nu_x}$, $p_{\nu_y}$ and the coefficients
$c_m$, $d_n$ are univariate polynomials of $p_{\nu_x}$ only. The resultant can then be obtained by
computing the determinant of the Sylvester matrix
\begin{eqnarray}
\mbox{Res}(p_{\nu_y}) & \!\! = \! & \! \mbox{Det} \! \left(
\begin{array}{cccc}
 c_0 &     & d_0 &  \\
 c_1 & c_0 & d_1 & d_0  \\
 c_2 & c_1 & d_2 & d_1  \\
     & c_2 &     & d_2  \\
\end{array}
\right) = 0 \;\;\;\;\;\;\; \\ \nonumber
\end{eqnarray}
which is equated to zero. The omitted elements of the matrix are identical to zero.
The resultant is a univariate polynomial of the form
\begin{eqnarray} \label{quartic}
 h_0p_{\nu_x}^4+h_1p_{\nu_x}^3+h_2p_{\nu_x}^2+h_3p_{\nu_x}+h_4
\end{eqnarray}
which contains the remaining unknown $p_{\nu_x}$. It is of degree four and 
can be solved analytically. The coefficients $h$ are given in the appendix. 
This result shows that there is at most a four fold ambiguity.
The neutrino and antineutrino masses are assumed to be zero in good approximation
in the following. They have been kept in the equations for the sake of completeness 
since the same set of equations can be exploited in search for new physics with the same decay topology
including invisible massive particles.
In Fig.~\ref{Nsol} the distribution of the number of solutions per event is plotted.
Here it has been assumed that the 4-vectors of the particles and the top quark and $W$ 
boson masses which enter into the system of equations are known exactly.
Under these conditions there are two solutions in about 80\% of cases and four solutions else.
In the next section it will be investigated how this distribution changes 
under more realistic conditions when the assumption
of exactness between particles and reconstructed objects is not valid anymore.
Once the solution of a neutrino momentum $p_{\nu_x}$ has been found 
the other neutrino and antineutrino
momentum components have to be determined. The antineutrino momentum $p_{\bar{\nu}_x}$
\begin{figure}[t]
\hspace*{0ex}\includegraphics[width=9.0cm]{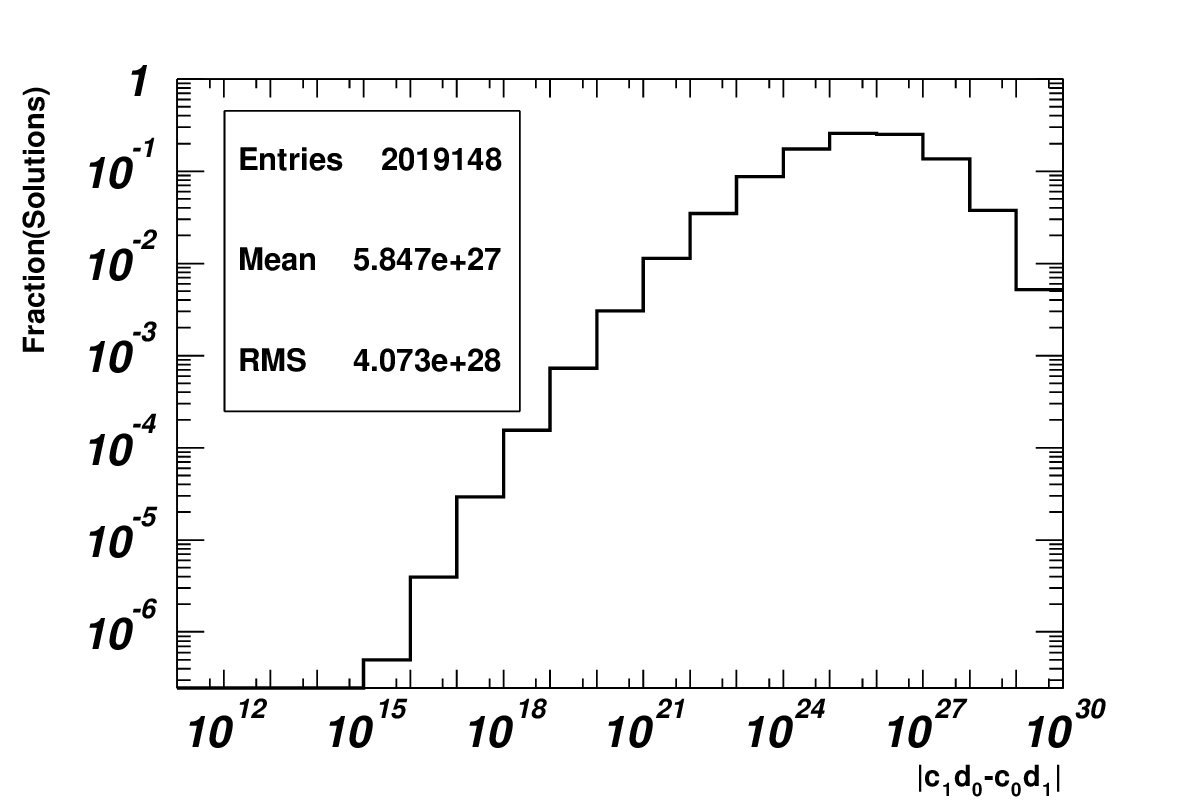}
\caption{\label{cd_diff_coeff} Distribution of the expression $c_1d_0-c_0d_1$ which appears in the
denominator in the solution of $p_{\nu_y}$.
Since the distribution is symmetric around zero the module of the expression is plotted. 
As can be seen the values assumed by the expression are far away from zero which would cause a 
singularity in the solution.}
\end{figure}
\begin{figure}[b]
\hspace*{0ex}\includegraphics[width=9.0cm]{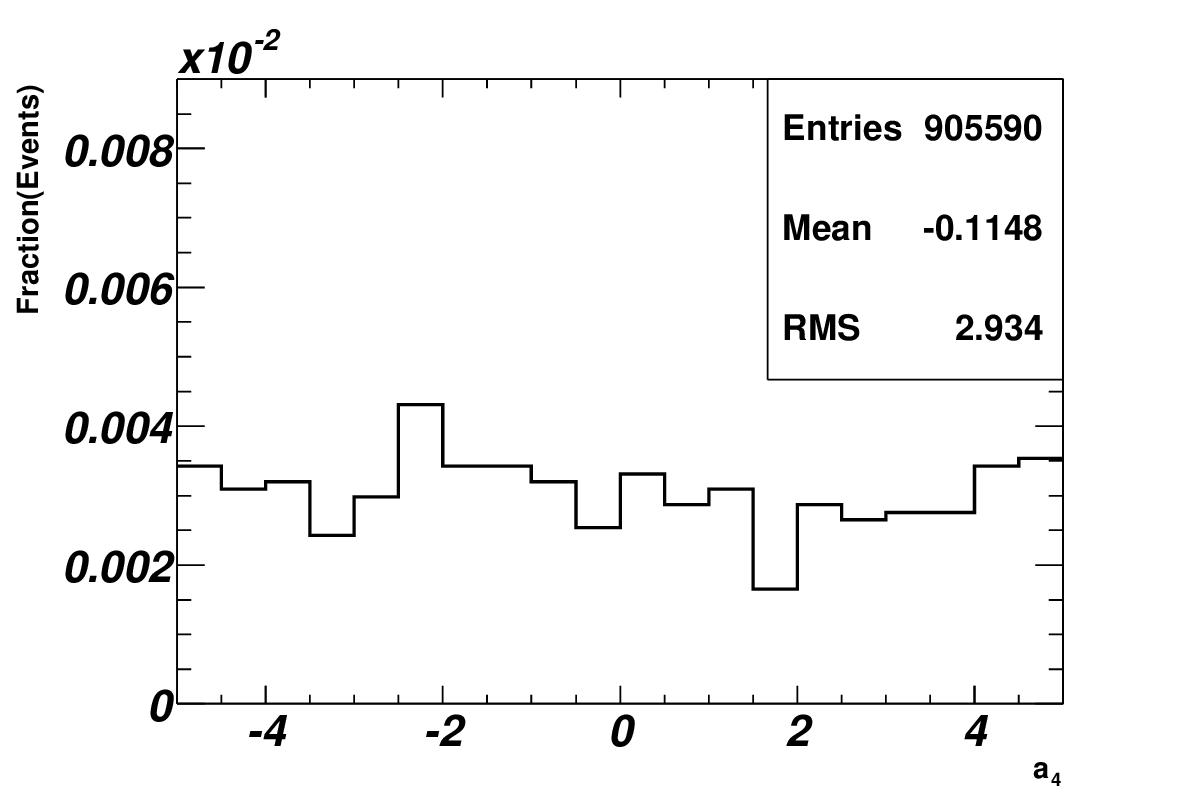}
\hspace*{0ex}\includegraphics[width=9.0cm]{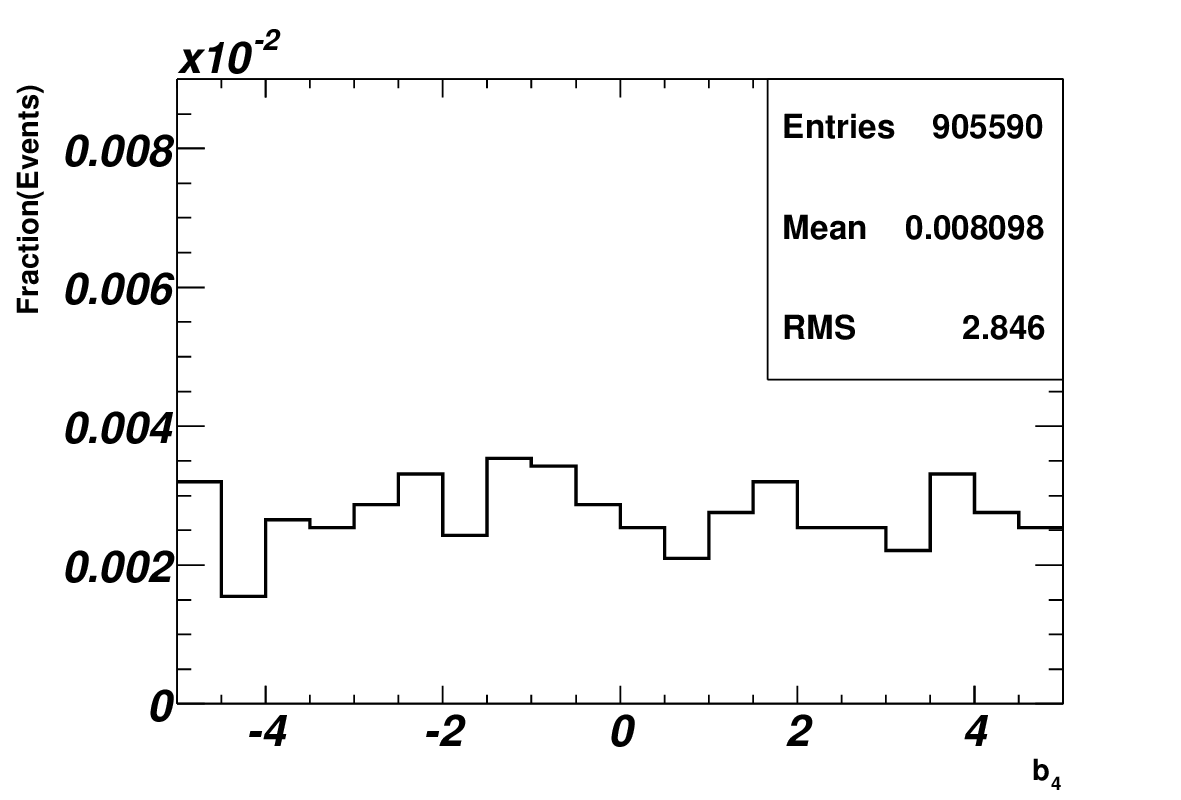}
\caption{\label{ab_coeff} Distributions of the coefficients $a_4$ (top) and $b_4$ (bottom).
The coefficients are flat distributed over the whole phase space including the value zero
where an irreducible singularity of their reciprocal resides.}
\end{figure}
can be immediately obtained by the linear transverse missing energy relation
of the initial system of equations (\ref{initialequations}).
To derive the neutrino momentum $p_{\nu_y}$ equation (\ref{cpoly}) is multiplied by $d_0$
and equation (\ref{dpoly}) is multiplied by $c_0$ so that their difference yields
a linear equation in the neutrino momentum $p_{\nu_y}$ which can then be isolated as
\begin{eqnarray} \label{pnuy}
 p_{\nu_y} & = & \frac{c_0d_2-c_2d_0}{c_1d_0-c_0d_1} .  
\end{eqnarray}
Again the antineutrino momentum $p_{\bar{\nu}_y}$
can be immediately obtained by the corresponding linear transverse missing energy relation
of the initial system of equations.
As shown in Fig.~\ref{cd_diff_coeff} the coefficient in the denominator of equation 
(\ref{pnuy}) does not acquire values
which are even close to the singularity at zero. Thus it is ensured that the neutrino momenta
$p_{\nu_y}$ and $p_{\bar{\nu}_y}$ can be computed accurately over the whole phase space of
possible solutions.

Finally the longitudinal (anti-)neutrino momenta $p_{\nu_z}$ and $p_{\bar{\nu}_z}$
can be easily obtained by the linear equations (\ref{linpnuz}) and (\ref{linpnubarz})
assuming that the coefficients $a_4$ and $b_4$ are different from zero since they appear 
as a product together with the longitudinal (anti-)neutrino momenta themselves. 
The distributions of the coefficients are shown in Fig.~\ref{ab_coeff}. The fraction of solutions 
close to the singularity - irreducible in the analytical solution - 
is below the per mill level 
and may be neglected for practical purposes. 
From a theoretical point of view this singularity can be circumvented in
solving the neutrino momenta $p_{\nu_z}$ and $p_{\bar{\nu}_z}$ analytically with 
the equations (2) and (3) of the algebraic approach \cite{ttdilepsol2005} which does 
not contain any singularity.
It has been verified that the longitudinal (anti-)neutrino momentum does not typically vanish
together with the coefficient $a_4$ ($b_4$) simultaneously, which would cause 
the singularity to disappear.

\begin{figure}[b]

\hspace*{0ex}\includegraphics[width=9.1cm]{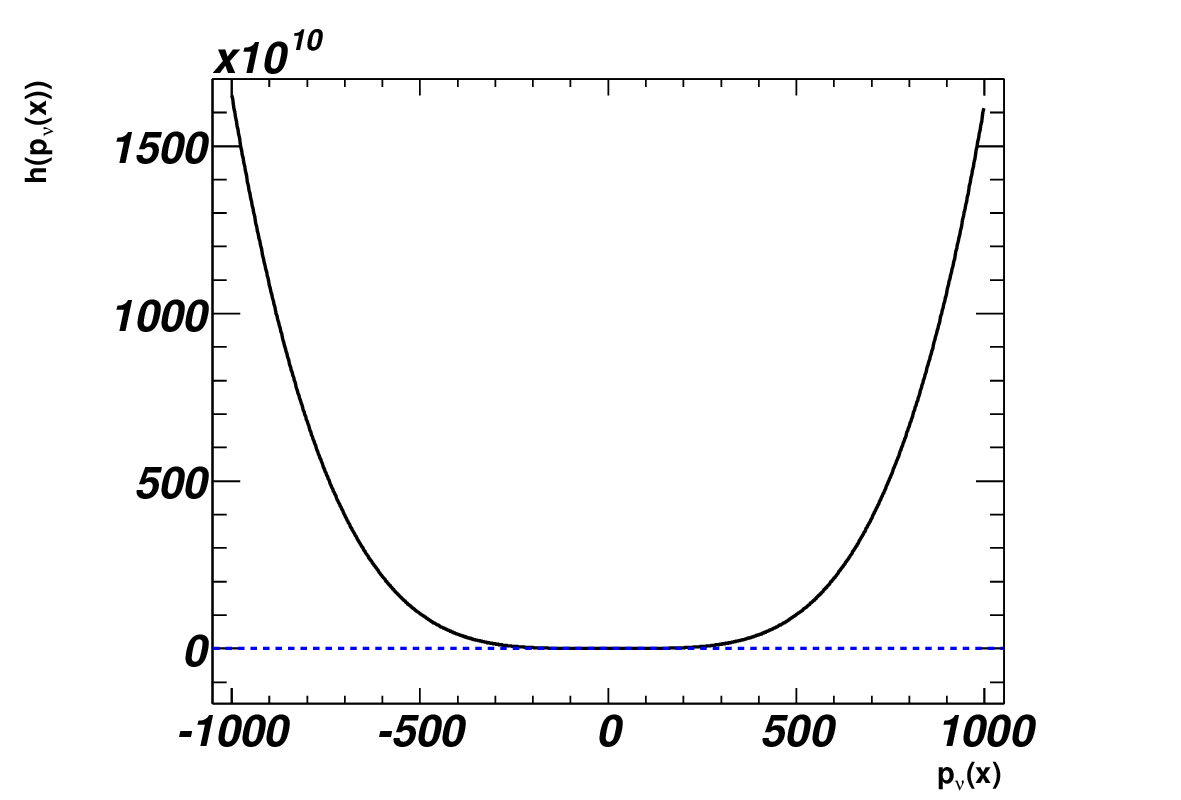}
\vspace*{-3.5ex}

\hspace*{0ex}\includegraphics[width=9.1cm]{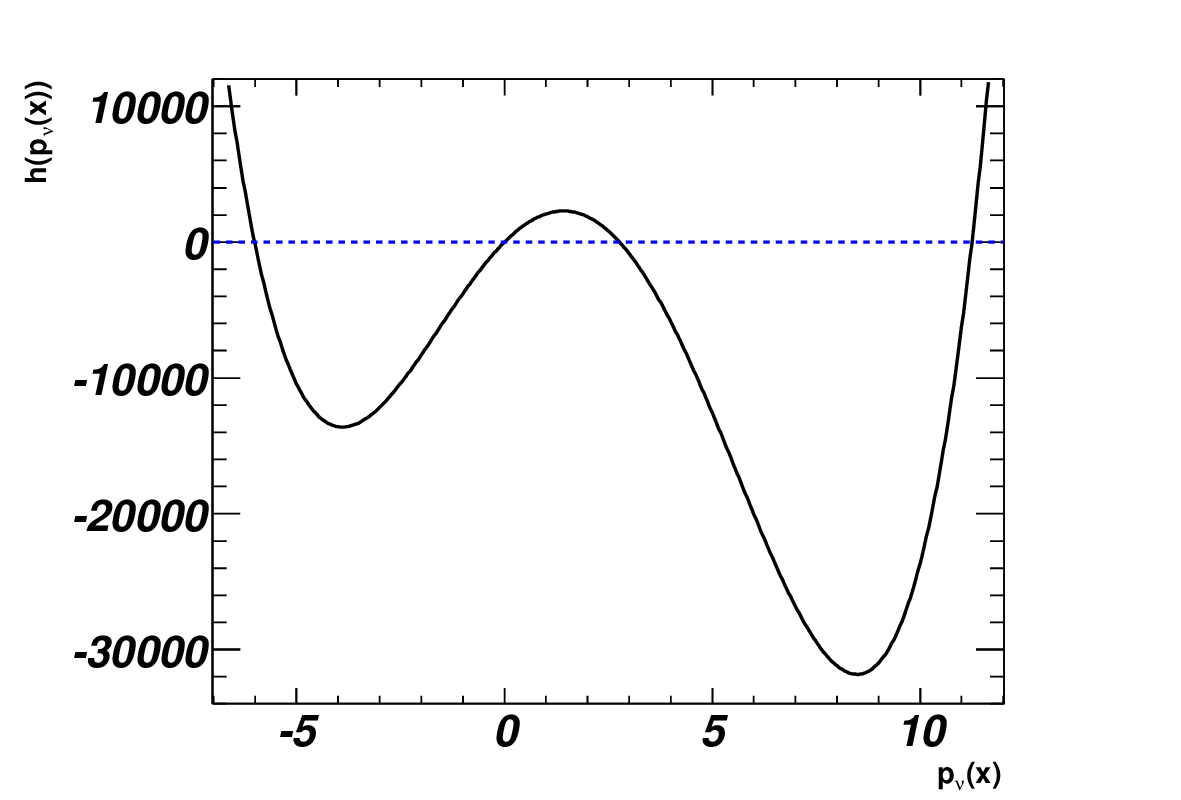}
\vspace*{-5ex}

\caption{\label{f4sol} A typical quartic equation whose real roots in $p_{\nu_x}$
are solutions of the initial system of equations describing the $t\bar{t}$ dilepton
kinematics. The bottom plot is zoomed around the interesting $p_{\nu_x}$
range of the abscissa where the analytical solution becomes singular.}
\end{figure}

\begin{table}[t]
\begin{tabular}{lccc} 
\hline
 & $\frac{N_{\mbox{\scriptsize sol}}=2}{N_{\mbox{\scriptsize sol}}>0}\;$ & $\;<\!N_{\mbox{\scriptsize sol}}^{>0}\!>\;$ & $\;$RMS($N_{\mbox{\scriptsize sol}}^{>0}$) \\ \hline  
$t, W$ masses known exactly & 0.82 & 2.37 & 0.77  \\ \hline
$W$ mass known exactly & & & \\ 
$t$ pole mass assumed & \raisebox{1.5ex}[-1.5ex]{0.84} & \raisebox{1.5ex}[-1.5ex]{2.32} & \raisebox{1.5ex}[-1.5ex]{0.74} \\ \hline 
$t,W$ pole mass assumed & 0.85 & 2.31 & 0.72  \\ \hline 
$t,W$ pole mass assumed & & &  \\ 
both $b \bar{b}$ permutations & \raisebox{1.5ex}[-1.5ex]{0.59} & \raisebox{1.5ex}[-1.5ex]{3.00} & \raisebox{1.5ex}[-1.5ex]{1.35} \\ \hline 
reconstructed $b$-jets & & & \\
(parton matched) & \raisebox{1.5ex}[-1.5ex]{0.79} & \raisebox{1.5ex}[-1.5ex]{2.42} & \raisebox{1.5ex}[-1.5ex]{0.82} \\ \hline
wrong $b$-jet permutation & & & \\
(parton matched) & \raisebox{1.5ex}[-1.5ex]{0.82} & \raisebox{1.5ex}[-1.5ex]{2.36} & \raisebox{1.5ex}[-1.5ex]{0.77} \\ \hline
both $b$-jet permutations & & & \\
(parton matched) & \raisebox{1.5ex}[-1.5ex]{0.52} & \raisebox{1.5ex}[-1.5ex]{3.22} & \raisebox{1.5ex}[-1.5ex]{1.48} \\ \hline
both $b$-jet permutations & & & \\
(parton matched, & 0.54 & 3.19 & 1.47  \\ 
 jets + leptons smeared) & & & \\ \hline
both $b$-jet permutations & & & \\
(parton matched, & & & \\ 
 jets + leptons smeared), & 0.0072 & 7.96 & 4.72 \\ 
reconstructed objects & & & \\
100 $\times$ resolution smeared & & & \\ \hline
\end{tabular}
\caption{\label{solution_characteristics2} Number of solutions, fractions and statistical quantities 
for events which have been solved ($N_{\mbox{\scriptsize sol}}>0$). The left column 
shows the fraction of events having exactly two solutions. 
In the centre the average number of solutions per solved event
is given. To the right the RMS of this distribution is shown.}
\end{table}

\section{Performance of the method}

\noindent
The performance studies discussed here are assuming Tevatron proton antiproton collider 
settings with a centre of mass energy of $1.96\,\mbox{TeV}$ which has been set up
in the Monte Carlo event generator PYTHIA 6.220 \cite{PYTHIA2001}. Cross-checks at a centre
of mass energy of $14\,\mbox{TeV}$ assuming the LHC proton collider environment 
confirm the independence of the method of particular collider settings.

The quartic equation in $p_{\nu_x}$ (\ref{quartic}) is typically flat around the neutrino momenta
of interest where solutions are to be expected. Fig.~\ref{f4sol} shows
the function for a given event. Only in the bottom plot where the
function is zoomed out the roots can be recognized.
Computationally the solutions are robust. 
The generated (anti-)neutrino momenta coincide with one of the solutions
to real precision assuming that the 4-vectors of the two leptons and the $b$, $\bar{b}$
quarks and the masses of the (anti-)top quarks and $W$ bosons which are 
entered into the quartic equation are known exactly.
The fraction of events where no solution can be found or no solution coincides
with the generated (anti-)neutrino momenta to real precision is below the per mill level.
If the $W$ boson mass is generated off-shell while its pole mass is assumed in the solution
the efficiency drops to 89\%. Relaxing the same assumption for the top quark mass results into
a further decrease of efficiency to 84\%. 
Beyond, an infrared-safe cone algorithm \cite{H11999} with cone size $R=0.5$ in the 
space spanned by pseudorapidity and azimuthal angle has been applied to the 
hadronic final state particles. Two reconstructed jets, two leptons and 
missing transverse energy are required
for an event to be selected. The jets are accepted as $b$-tagged if they coincide within 
$\Delta R<0.5$ with the generated $b$ quarks.
The solution efficiency drops to 71\% and can be re-established at 81\% in solving both
$b$ quark jet permutations.
Smearing the leptons and jets with the energy resolution of the D0 detector \cite{D0note4677}
decreases the efficiency to 75\%.
In practice, a given event 
can be solved repeatedly, with the energy of the particles and objects smeared randomly within 
the detector resolution once each iteration to improve the solution efficiency.
These observations are consistent with the
findings of the algebraic approach \cite{ttdilepsol2005}. This confirms on one hand the 
reliability of the algebraic approach and rises on the other hand the question 
what numerical methods with a superior solution efficiency are actually solving.

Considering only events which could be solved it is important to investigate
the number of solutions in dependence of the experimental settings since
this number is directly proportional to the ambiguities of the solved and 
reconstructed events which in turn determines the significance of the solutions
and any observable making use of it.
In Tab.~\ref{solution_characteristics2} the fraction of solved events having exactly 
two solutions, the average number of solutions and its RMS is given for different
experimental settings. The first four lines describe the evolution of these quantities 
derived from the particle final state. Relaxing the amount of assumption 
about the top quark and $W$ boson masses increases the fraction of solved events 
with exactly two solutions while the average number of solutions and its RMS decrease
slightly. Allowing both $b$ quark jet permutations - assuming that the charge of the
quarks can not be determined with adequate certainty - the fraction of events
having exactly two solutions drops considerably in favour of a higher solution multiplicity
with larger RMS.
The table items below show the number of solutions for reconstructed objects, first for right, 
wrong and both $b$ quark jet permutations then energy resolution smearing is applied to the 
reconstructed objects and finally 100 solution iterations have been 
accomplished to take into account the
uncertainty in the measured energy of the reconstructed objects. The general tendency 
is that the fraction of solved events with exactly two solutions decreases with less accurate
knowledge about the particles and objects while the solution multiplicity and its RMS does 
increase.

\section{Conclusions}

\noindent
The analytical solution of the system of equations describing the $t\bar{t}$
dilepton kinematics has been presented. The Ansatz of formulating two equations
linear in the three neutrino and antineutrino momentum components
leads after substitution of the longitudinal (anti-)neutrino momenta to two 
multivariate polynomials of two unknowns with multidegree two.
It turns out that each of these two polynomials has a singularity which can be removed.
In contrast there are two irreducible singularities in the linear equations described 
above which can be circumvented in exploiting the analytical Ansatz of the algebraic 
approach \cite{ttdilepsol2005} to determine the longitudinal (anti-)neutrino momenta.
The two multivariate polynomials can be reduced to a univariate polynomial 
of degree four by means of resultants.
The obtained quartic equation is solved analytically.
The solution could be derived without any use of computer algebra software.
The fraction of events without any solution or with no solution matching the generated
(anti-)neutrino momenta with real precision are below the per mill 
level assuming that the particle momenta and masses inserted into the analytical 
solution are known exactly.
Consistent with the observations of the algebraic approach \cite{ttdilepsol2005}
little deviations of the inserted particle momenta and masses from their true
values drop the solution efficiency and purity considerably.
At the same time the solution multiplicity increases.
This raises the question what more efficient numerical methods are actually solving.
General solution methods can compare their performance
with the analytical solution described here.

\begin{acknowledgments}

\noindent 
Many thanks to Vladislav Simak and Petr Homola for their support and many useful discussions.
I'm also grateful for helpful suggestions from colleagues of LPNHE, Universit\'es de Paris VI, VII
and the D\O\ collaboration.
This work has been supported by the 
{\it Commissariat \`a l'Energie Atomique} and
CNRS/{\it{Institut National de Physique Nucl\'eaire et de Physique des
Particules}}, France.
Many thanks to Bruno Wittmer and Georgios Anagnostou for useful discussions.

\end{acknowledgments}

\appendix

\section*{Appendix}

\subsection*{Polynomial coefficients}

\noindent
The coefficients of equation (\ref{linpnuz}) are given by
\begin{eqnarray} \nonumber
a_1 & = & (E_b+E_{\ell^+})(m_W^2-m_{\ell^+}^2-m_{\nu}^2) \\ \nonumber
    &   & -E_{\ell^+}(m_t^2-m_b^2-m_{\ell^+}^2-m_{\nu}^2) \\ \nonumber
    &   & +2E_bE_{\ell^+}^2-2E_{\ell^+}\vec{p_b}\vec{p_{\ell^+}} , \\ \nonumber
a_2 & = & 2(E_bp_{\ell^+_x}-E_{\ell^+}p_{b_x}) , \\ \nonumber
a_3 & = & 2(E_bp_{\ell^+_y}-E_{\ell^+}p_{b_y}) , \\ \nonumber
a_4 & = & 2(E_bp_{\ell^+_z}-E_{\ell^+}p_{b_z}) 
\end{eqnarray}
where it is important that the coefficient $a_4$ does not vanish since equation
(\ref{linpnuz}) has to be divided by it to isolate the unknown $p_{\nu_z}$.
As explained in section \ref{anasol} this irreducible singularity can be circumvented in solving for
$p_{\nu_z}$ with the analytical Ansatz made in the algebraic approach \cite{ttdilepsol2005}.

The equivalent equation of the antitop quark parton branch is
\begin{eqnarray}
  0 & = & b_1 + b_2p_{\bar{\nu}_x} + b_3p_{\bar{\nu}_y} + b_4p_{\bar{\nu}_z}
\end{eqnarray} 
with the coefficients
\begin{eqnarray} \nonumber
b_1 & = & (E_{\bar{b}}+E_{\ell^-})(m_W^2-m_{\ell^-}^2-m_{\bar{\nu}}^2) \\ \nonumber
    &   & -E_{\ell^-}(m_t^2-m_{\bar{b}}^2-m_{\ell^-}^2-m_{\bar{\nu}}^2) \\ \nonumber
    &   & +2E_{\bar{b}}E_{\ell^-}^2-2E_{\ell^-}\vec{p_{\,\overline{\!b}}}\vec{p_{\ell^-}} , \\ \nonumber
b_2 & = & 2(E_{\bar{b}}p_{\ell^-_x}-E_{\ell^-}p_{\bar{b}_x}) , \\ \nonumber
b_3 & = & 2(E_{\bar{b}}p_{\ell^-_y}-E_{\ell^-}p_{\bar{b}_y}) , \\ \nonumber
b_4 & = & 2(E_{\bar{b}}p_{\ell^-_z}-E_{\ell^-}p_{\bar{b}_z}) .
\end{eqnarray}
Again there is a singularity in case of vanishing coefficient $b_4$.
The coefficients of equation (\ref{pnuxypol}) are given by
\begin{eqnarray} \nonumber
 c_{22} & = & (m_{W^+}^2-m_{\ell^+}^2-m_{\nu}^2)^2 - 4({E_{\ell^+}}^2-p_{\ell^+_z}^2)(a_1/a_4)^2 \\ \nonumber
     &   & -4(m_{W^+}^2-m_{\ell^+}^2-m_{\nu}^2)p_{\ell^+_z}a_1/a_4 , \\ \nonumber
 c_{21} & = & 4(m_{W^+}^2-m_{\ell^+}^2-m_{\nu}^2)(p_{\ell^+_x}-p_{\ell^+_z}a_2/a_4) \\ \nonumber
        & & -8(E_{\ell^+}^2-p_{\ell^+_z}^2)a_1a_2/a_4^2 
            -8p_{\ell^+_x}p_{\ell^+_z}a_1/a_4  , \\ \nonumber
 c_{20} & = & -4(E_{\ell^+}^2-p_{\ell^+_x}^2)
              -4(E_{\ell^+}^2-p_{\ell^+_z}^2)(a_2/a_4)^2 \\ \nonumber
        & & -8p_{\ell^+_x}p_{\ell^+_z}a_2/a_4  , \\ \nonumber
 c_{11} & = & 4(m_{W^+}^2-m_{\ell^+}^2-m_{\nu}^2)(p_{\ell^+_y}-p_{\ell^+_z}a_3/a_4) \\ \nonumber
   & & -8(E_{\ell^+}^2-p_{\ell^+_z}^2)a_1a_3/a_4^2
       -8p_{\ell^+_y}p_{\ell^+_z}a_1/a_4 , \\ \nonumber
 c_{10} & = & -8(E_{\ell^+}^2-p_{\ell^+_z}^2)a_2a_3/a_4^2+8p_{\ell^+_x}p_{\ell^+_y} \\ \nonumber
        & & -8p_{\ell^+_x}p_{\ell^+_z}a_3/a_4 -8p_{\ell^+_y}p_{\ell^+_z}a_2/a_4 , \\ \nonumber
 c_{00} & = & -4(E_{\ell^+}^2-p_{\ell^+_y}^2)-4(E_{\ell^+}^2-p_{\ell^+_z}^2)(a_3/a_4)^2 \\ \nonumber
        & &    -8p_{\ell^+_y}p_{\ell^+_z}a_3/a_4 .
\end{eqnarray}
Similar the coefficients $d'$ of the antitop quark branch depend on the
coefficients $b$ in the following way
\begin{eqnarray} \nonumber
  d'_{22} & = & (m_{W^-}^2-m_{\ell^-}^2-m_{\bar{\nu}}^2)^2-4(E_{\ell^-}^2-p_{\ell^-_z}^2)(b_1/b_4)^2 \\ \nonumber
    & & -4(m_{W^-}^2-m_{\ell^-}^2-m_{\bar{\nu}}^2)p_{\ell^-_z}b_1/b_4 ,  \\ \nonumber
  d'_{21} & = & 4(m_{W^-}^2-m_{\ell^-}^2-m_{\bar{\nu}}^2)(p_{\ell^-_x}-p_{\ell^-_z}b_2/b_4) \\ \nonumber
          & & -8(E_{\ell^-}^2-p_{\ell^-_z}^2)b_1b_2/b_4^2 -8p_{\ell^-_x}p_{\ell^-_z}b_1/b_4 , \\ \nonumber
  d'_{20} & = & -4(E_{\ell^-}^2-p_{\ell^-_x}^2) -4(E_{\ell^-}^2-p_{\ell^-_z}^2)(b_2/b_4)^2 \\ \nonumber
          & & -8p_{\ell^-_x}p_{\ell^-_z}b_2/b_4 , \\ \nonumber
  d'_{11} & = & 4(m_{W^-}^2-m_{\ell^-}^2-m_{\bar{\nu}}^2)(p_{\ell^-_y}-p_{\ell^-_z}b_3/b_4) \\ \nonumber
          & & -8(E_{\ell^-}^2-p_{\ell^-_z}^2)b_1b_3/b_4^2 -8p_{\ell^-_y}p_{\ell^-_z}b_1/b_4  , \\ \nonumber
  d'_{10} & = & -8(E_{\ell^-}^2-p_{\ell^-_z}^2)b_2b_3/b_4^2 +8p_{\ell^-_x}p_{\ell^-_y} \\ \nonumber 
          & & -8p_{\ell^-_x}p_{\ell^-_z}b_3/b_4 -8p_{\ell^-_y}p_{\ell^-_z}b_2/b_4 , \\ \nonumber
  d'_{00} & = & -4(E_{\ell^-}^2-p_{\ell^-_y}^2) -4(E_{\ell^-}^2-p_{\ell^-_z}^2)(b_3/b_4)^2 \\ \nonumber
          & & -8p_{\ell^-_y}p_{\ell^-_z}b_3/b_4 \; .
\end{eqnarray}
The remaining unknowns in these equations - which are the transverse antineutrino momenta - 
are substituted by the missing transverse energy relations of the system of equations (\ref{initialequations})  
to obtain finally the set of equations
\begin{eqnarray} \nonumber
  d_{22} & = & d'_{22}+E_x\!\!\!\!\!\!/ \;\;^2 d'_{20} +E_y\!\!\!\!\!\!/ \;\;^2 d'_{00}
              +E_x\!\!\!\!\!\!/ \;\; E_y\!\!\!\!\!\!/ \;\; d'_{10} \\ \nonumber
         & & +E_x\!\!\!\!\!\!/ \;\; d'_{21} +E_y\!\!\!\!\!\!/ \;\; d'_{11} , \\ \nonumber
  d_{21} & = & -d'_{21} -2E_x\!\!\!\!\!\!/ \;\; d'_{20} -E_y\!\!\!\!\!\!/ \;\; d'_{10} , \\ \nonumber
  d_{20} & = & d'_{20} , \\ \nonumber
  d_{11} & = & -d'_{11} -2E_y\!\!\!\!\!\!/ \;\; d'_{00} -E_x\!\!\!\!\!\!/ \;\; d'_{10} \\ \nonumber
  d_{10} & = & d'_{10} , \\ \nonumber
  d_{00} & = & d'_{00} , \\ \nonumber
\end{eqnarray}
which depends merely on the transverse neutrino momenta $p_{\nu_x}$ and $p_{\nu_y}$.

The resultant expressed in terms of the multivariate polynomials $c_{jk}$ and $d_{mn}$
are given by
\begin{eqnarray} \nonumber
  h_4 & = & c_{00}^2d_{22}^2+c_{11}d_{22}(c_{11}d_{00}-c_{00}d_{11}) \\ \nonumber
     & &   +c_{00}c_{22}(d_{11}^2\!-\!2d_{00}d_{22})+c_{22}d_{00}(c_{22}d_{00}\!-\!c_{11}d_{11}) , \\ \nonumber 
  h_3 & = & c_{00}d_{21}(2c_{00}d_{22}\!-\!c_{11}d_{11})+c_{00}d_{11}(2c_{22}d_{10}\!+\!c_{21}d_{11}) \\ \nonumber
       & & +c_{22}d_{00}(2c_{21}d_{00}\!-\!c_{11}d_{10})-c_{00}d_{22}(c_{11}d_{10}\!+\!c_{10}d_{11}) \\ \nonumber 
       & & -2c_{00}d_{00}(c_{22}d_{21}\!+\!c_{21}d_{22})-d_{00}d_{11}(c_{11}c_{21}\!+\!c_{10}c_{22}) \\ \nonumber
       & & +c_{11}d_{00}(c_{11}d_{21}+2c_{10}d_{22}) , \\ \nonumber 
%
  h_2 & = & c_{00}^2(2d_{22}d_{20}+d_{21}^2)-c_{00}d_{21}(c_{11}d_{10}+c_{10}d_{11}) \\ \nonumber
       & & +c_{11}d_{20}(c_{11}d_{00}-c_{00}d_{11})+c_{00}d_{10}(c_{22}d_{10}-c_{10}d_{22})  \\ \nonumber
       & & +c_{00}d_{11}(2c_{21}d_{10}+c_{20}d_{11})+(2c_{22}c_{20}+c_{21}^2)d_{00}^2  \\ \nonumber
       & & -2c_{00}d_{00}(c_{22}d_{20}+c_{21}d_{21}+c_{20}d_{22})  \\ \nonumber 
       & & +c_{10}d_{00}(2c_{11}d_{21}+c_{10}d_{22})-d_{00}d_{10}(c_{11}c_{21}+c_{10}c_{22}) \\ \nonumber 
       & & -d_{00}d_{11}(c_{11}c_{20}+c_{10}c_{21}) ,  \\ \nonumber 
  h_1 & = & c_{00}d_{21}(2c_{00}d_{20}\!-\!c_{10}d_{10})-c_{00}d_{20}(c_{11}d_{10}\!+\!c_{10}d_{11}) \\ \nonumber
       & & +c_{00}d_{10}(c_{21}d_{10}\!+\!2c_{20}d_{11})\!-\!2c_{00}d_{00}(c_{21}d_{20}\!+\!c_{20}d_{21}) \\ \nonumber
       & & +c_{10}d_{00}(2c_{11}d_{20}\!+\!c_{10}d_{21})\!-\!c_{20}d_{00}(2c_{21}d_{00}\!-\!c_{10}d_{11}) \\ \nonumber
       & & -d_{00}d_{10}(c_{11}c_{20}+c_{10}c_{21}) , \\ \nonumber 
  h_0 & = & c_{00}^2d_{20}^2+c_{10}d_{20}(c_{10}d_{00}-c_{00}d_{10}) \\ \nonumber
      & & +c_{20}d_{10}(c_{00}d_{10}\!-\!c_{10}d_{00})+c_{20}d_{00}(c_{20}d_{00}\!-\!2c_{00}d_{20}) . 
\end{eqnarray}

To avoid singularities which arise in the case of vanishing factors $a_4$ or $b_4$ the
coefficients $c_{jk}$, $d_{mn}$ of the polynomials (\ref{pnuxypol}) and (\ref{pnubarxypol}) 
have been multiplied with the least common multiple of the denominators which are
$a_4^2$ and $b_4^2$ respectively. These factors are constant for a given event 
and thus do not alter the position of the real roots which 
correspond to the neutrino momenta $p_{\nu_x}$.

\subsection*{Quartic equation}

The quartic equation can be solved analytically in reducing it to a cubic equation.
There are several ways to achieve this. Here the method of 
Ferrari \cite{Ferrari} - who was the first to develop an algebraic technique 
for solving the general quartic equation - is being used. 

First the leading coefficient $h_0$ of the 
quartic polynomial (\ref{quartic}) is normalized to one 
(in the case the leading coefficient vanishes
the problem is already reduced to a cubic equation).
If the constant $h_4$ vanishes the quartic polynomial can be factorized
into $p_{\nu_x}$ times a cubic equation. In this case one root namely $p_{\nu_x}=0$
is already known.
The substitution $p_{\nu_x} = p'_{\nu_x}-h_1/4$ leads to the simplified
equation
\begin{eqnarray} \nonumber
  0 & = & p'^4_{\nu_x} +k_1p'^2_{\nu_x} +k_2p'_{\nu_x}+k_3
\end{eqnarray}
with the coefficients
\begin{eqnarray} \nonumber
 k_1 & = & h_2-3h_1^2/8 , \\ \nonumber
 k_2 & = & h_3+h_1^3/8-h_1h_2/2 , \\ \nonumber
 k_3 & = & h_4-3h_1^4/256+h_1^2h_2/16-h_1h_3/4 .
\end{eqnarray}
If the coefficient $k_3$ vanishes again the equation can be factorized into
$p'_{\nu_x}$ times a cubic polynomial.
If the coefficient $k_2$ vanishes the quartic polynomial in $p'_{\nu_x}$
can be expressed as a quadratic equation in ${p'^2_{\nu_x}}$.
In the general case where all three coefficients $k_1$, $k_2$ and $k_3$
are different from zero the quartic polynomial can be factorized into
the product of two quadratic polynomials 
as follows
\begin{eqnarray} \label{quarticfactorized}
  p'^4_{\nu_x} +k_1p'^2_{\nu_x} +k_2p'_{\nu_x}+k_3 = \\ \nonumber
  (p'^2_{\nu_x}+t_1p'_{\nu_x}+t_2) (p'^2_{\nu_x}-t_1p'_{\nu_x}+k_3/t_2) .
\end{eqnarray}
Once the new coefficients $t_1$ and $t_2$ have been determined the quadratic polynomials
can be easily solved.
Comparison of the coefficients yields
\begin{eqnarray} \nonumber
  k_1 & = & k_3/t_2+t_2-t_1^2 
\end{eqnarray}
and
\begin{eqnarray} \nonumber
  k_2 & = & t_1 (k_3/t_2-t_2) .
\end{eqnarray}
It is ensured that $t_2$ which appears in the denominator does not vanish 
since the coefficient $k_3$ has been assumed to be different from zero and $k_1$, $k_2$
are finite. Eliminating $t_2$ in the two nonlinear equations above
leads to a cubic equation in $t_1^2$.
To achieve this the two equations above are rewritten in the following form
\begin{eqnarray} \nonumber
  k_3/t_2+t_2 & = & k_1+t_1^2 , \\ \nonumber 
  k_3/t_2-t_2 & = & k_2/t_1 .
\end{eqnarray}
Adding and subtracting them leads to
\begin{eqnarray}
   2k_3/t_2 & = & k_1+t_1^2+k_2/t_1 , \\ 
   2t_2     & = & k_1+t_1^2-k_2/t_1 \label{coeffcomp}
\end{eqnarray}
whose product can finally be written as
\begin{eqnarray} \nonumber
  0 & = & t_1^6+2k_1t_1^4+(k_1^2-4k_3)t_1^2-k_2^2 
\end{eqnarray}
which is a cubic equation in $t_1^2$. Any positive root of $t_1^2$
can be used to derive all real solutions of the initial quartic equation
(negative roots would lead to imaginary values of $\pm t_1$. Either sign
can be used to solve the factorized quartic equation. Changing the sign corresponds 
to swapping the coefficients between the first and the second quadratic polynomial
in equation (\ref{quarticfactorized}).
Descartes' Sign Rule \cite{Descartes} can be exploited to ensure that there
is always at least one positive root. According to the rule the number of
sign changes of the consecutive polynomial coefficients is the maximal number of positive
roots. Now one can substitute $t_1^2$ by $-t_1^2$ to determine
the maximal number of negative roots.  
Since $k_2$ is real the constant coefficient 
$-k_2^2$ is negative. The leading monomial has also a negative coefficient.
Thus there can be two or zero sign changes. A cubic equation with real coefficients 
has always either one or three real roots. 
In the case of two or zero negative roots there must 
conclusively be at least one positive root.
Once this root has been determined, $t_1$ can be inserted into equation 
(\ref{coeffcomp}) above to determine $t_2$ and subsequently the quadratic 
polynomials (\ref{quarticfactorized}) of the quartic equation.

\subsection*{Cubic equation}

There are several ways to solve the cubic equation \cite{Tartaglia}. Here the approach  
of \cite{Press1994} has been adopted. The cubic equation
\begin{eqnarray} \nonumber
  0 & = & z^3 + s_1z^2 +s_2z +s_3 
\end{eqnarray}
is assumed to have real coefficients.
First the two variables
\begin{eqnarray} \nonumber
  q = \frac{s_1^2-3s_2}{9} 
\end{eqnarray}
and
\begin{eqnarray} \nonumber
  r = \frac{2s_1^3-9s_1s_2+27s_3}{54} 
\end{eqnarray}
are determined. If $r^2 < q^3$ the cubic equation has three real roots which can be found
by computing
\begin{eqnarray} \nonumber
  \theta = \arccos{r/\sqrt{q^3}} . 
\end{eqnarray}
The three roots are then given by
\begin{eqnarray} \nonumber
  z_1 & = & -2\sqrt{q}\cos(\frac{\theta}{3})-\frac{s_1}{3} , \\ \nonumber
  z_1 & = & -2\sqrt{q}\cos(\frac{\theta+2\pi}{3})-\frac{s_1}{3} , \\ \nonumber
  z_1 & = & -2\sqrt{q}\cos(\frac{\theta-2\pi}{3})-\frac{s_1}{3} . 
\end{eqnarray}
Their first appearance goes back to Fran\c{c}ois Vi\`{e}ta who published them in 1615.
In the case of $r^2 \geq q^3$ there is only one real solution and defining the auxiliary variables
\begin{eqnarray} \nonumber
  u = (-r+\sqrt{r^2-q^3})^{1/3} 
\end{eqnarray}
and
\begin{eqnarray} \nonumber
  v = (-r-\sqrt{r^2-q^3})^{1/3} 
\end{eqnarray}
allows to express the real solution simply in terms of $u$ and $v$ as
\begin{eqnarray} \nonumber
  z_1 = u + v - \frac{s_1}{3} .
\end{eqnarray}

\bibliography{apssamp}

\begin{thebibliography}{3}
\expandafter\ifx\csname natexlab\endcsname\relax\def\natexlab#1{#1}\fi
\expandafter\ifx\csname bibnamefont\endcsname\relax
  \def\bibnamefont#1{#1}\fi
\expandafter\ifx\csname bibfnamefont\endcsname\relax
  \def\bibfnamefont#1{#1}\fi
\expandafter\ifx\csname citenamefont\endcsname\relax
  \def\citenamefont#1{#1}\fi
\expandafter\ifx\csname url\endcsname\relax
  \def\url#1{\texttt{#1}}\fi
\expandafter\ifx\csname urlprefix\endcsname\relax\def\urlprefix{URL }\fi
\providecommand{\bibinfo}[2]{#2}
\providecommand{\eprint}[2][]{\url{#2}}






\bibitem[{\citenamefont{Dalitz}()}]{Dalitz1992}
\bibinfo{author}{\bibnamefont{R. H. Dalitz, G. R. Goldstein}},
{\sl ``Analysis of top-antitop production and dilepton decay events and the top quark mass''}, 
{Phys. Lett. {\bf B} 287, 225 (\bibinfo{year}{1992})}

\bibitem[{\citenamefont{Goldstein}()}]{Goldstein1992}
\bibinfo{author}{\bibnamefont{R. H. Dalitz, G. R. Goldstein}},
{\sl ``Decay and polarization properties of the top quark''}, 
{Phys. Rev. {\bf D45}, 1531 (\bibinfo{year}{1992})}

\bibitem[{\citenamefont{Homola}()}]{Homola2004}
\bibinfo{author}{\bibnamefont{I. Borjanovi\'c {\it et al}}},
{\sl ``Investigation of the top mass measurements with the ATLAS detector at LHC''}, 
hep-ex/0403021, \bibinfo{year}{2004}


\bibitem[{\citenamefont{Maple}()}]{Maple2004}
\bibinfo{author}{\bibfnamefont{Waterloo Maple Inc., Ontario, Canada}},
Maple,
{\tt http://www.maplesoft.com}, 
\bibinfo{year}{2004}.
 

\bibitem[{\citenamefont{Homola2}()}]{Homola2005}
\bibinfo{author}{\bibnamefont{P. Homola}},
private communication, \bibinfo{year}{2005}


\bibitem[{\citenamefont{ttdilepsol}()}]{ttdilepsol2005}
\bibinfo{author}{\bibnamefont{L. Sonnenschein}},
{\sl ``Algebraic approach to solve $t\bar{t}$ dilepton equations''}, 
{Physical Rev. \bf D \rm 72 (\bibinfo{year}{2005}) 095020.}






\bibitem[{\citenamefont{Zhou}()}]{Zhou1998}
\bibinfo{author}{\bibfnamefont{H.-Y. Zhou.}}, 
{\sl ``CP violation in top quark pair production at hadron colliders''}, 
{Physical Rev. \bf D \rm 58 (\bibinfo{year}{1998}) 114002.} 


\bibitem[{\citenamefont{Sjolin}()}]{Sjolin2003}
\bibinfo{author}{\bibfnamefont{J. Sjolin}},
{\sl ``LHC experimental sensitivity to CP violating $gt\bar{t}$ couplings''}, 
{J. Phys. G (\bibinfo{year}{2003}): Nucl. Part. Phys. 29 543-560}



\bibitem[{\citenamefont{PYTHIA}()}]{PYTHIA2001}
\bibinfo{author}{\bibfnamefont{T. Sj\"ostrand, P. Eden, C. Friberg, L. L\"onnblad, G. Miu, S. Mrenna and E. Norrbin}}, 
{Computer Physics Commun. 135 (\bibinfo{year}{2001}) 238.}

\bibitem[{\citenamefont{H1}()}]{H11999}
\bibinfo{author}{\bibnamefont{H1 Collab., C. Adloff {\it et al}}},
{\sl ``Measurement of internal jet structure in dijet production in deep-inelastic scattering at HERA''}, 
{Nuclear Physics \bf B \rm 545 (\bibinfo{year}{1999}) 3-20.}

\bibitem[{\citenamefont{D0}()}]{D0note4677}
\bibinfo{author}{\bibnamefont{S. N. Fatakia, U. Heintz, L. Sonnenschein}},
{\sl ``Top Mass Measurement in the Dilepton Channel''}, 
{D$\emptyset$ note 4677}






\bibitem[{\citenamefont{Ferrari}()}]{Ferrari}
\bibinfo{author}{\bibfnamefont{Eric W. Weisstein.}}, 
{{\sl ``Quartic Equation''}, From MathWorld--A Wolfram Web Resource}, 
\url{http://mathworld.wolfram.com/QuarticEquation.html},
\bibinfo{year}{1999} 

\bibitem[{\citenamefont{Descartes}()}]{Descartes}
\bibinfo{author}{\bibfnamefont{Eric W. Weisstein.}}, 
{{\sl ``Descartes' Sign Rule''}, From MathWorld--A Wolfram Web Resource}, 
\url{http://mathworld.wolfram.com/DescartesSignRule.html},
\bibinfo{year}{1999} 

\bibitem[{\citenamefont{Tartaglia}()}]{Tartaglia}
\bibinfo{author}{\bibfnamefont{Eric W. Weisstein.}}, 
{{\sl ``Cubic Equation''}, From MathWorld--A Wolfram Web Resource}, 
\url{http://mathworld.wolfram.com/CubicFormula.html},
\bibinfo{year}{1999} 


\bibitem[{\citenamefont{Press}()}]{Press1994}
\bibinfo{author}{\bibfnamefont{W. H. Press {\it et al}}}, 
{{\sl ``Numerical Recipes in FORTRAN''}, Cambridge University Press},
\bibinfo{year}{1994} 





\end{thebibliography}

\pagebreak

\end{document}